\documentclass[american,english,twocolumn]{epl}
\usepackage[T1]{fontenc}
\usepackage[latin9]{inputenc}
\usepackage{amstext}
\usepackage{graphicx}
\usepackage{amssymb}

\makeatletter
\title{Thermal conductivity of mass-graded graphene flakes}

\author{J. Cheh\inst{1} \and H. Zhao\inst{1,2}}

\institute{                    
  \inst{1} Department of Physics, Institute of Theoretical Physics and Astrophysics, Xiamen University, Xiamen 361005, China \\
  \inst{2} Sate Key Laboratory for Nonlinear Mechanics, Institute of Mechanics, Chinese Academy of Sciences, Beijing 100080, China
}

\pacs{65.80.Ck}{Thermal properties of graphene}
\pacs{66.70.-f}{Nonelectronic thermal conduction and heat-pulse propagation in solids}
\pacs{81.05.Ue}{Graphene}

\makeatother

\usepackage{babel}

\makeatother

\usepackage{babel}

\begin{document}
\begin{abstract}
In this letter we study thermal conduction in mass-graded graphene
flakes by nonequilibrium molecular dynamics simulations. It is found
that mass-graded graphene flakes reveal no thermal rectification effect
in thermal conduction process. The dependence of thermal conductivity
upon the heat flux and the mass gradient are studied to confirm the
generality of the result.The mechanism leading to the absence of thermal
rectification effect is also discussed. 
\end{abstract}
Graphene, a single layer of carbon atoms in a honeycomb lattice with
sp$^{2}$ bonds, has attracted much interest due to its novel properties\cite{01.graphene01,02.graphene02}.
It reveals superior high thermal conductivity up to 2500-5000 W/mK
at room temperature\cite{03.baladin,04.cai}. Thus it has been considered
as a promising candidate for various kinds of thermal devices, such
as thermal rectifiers. Thermal rectification is a phenomenon that
the heat flux runs preferentially along one direction and inferiorly
along the opposite direction\cite{05.rect01,06.rect02,07.rect04,08.rect05.LIbaowen}.
Researchers have proposed several different graphene based thermal
rectifiers by introducing asymmetric shapes in molecular dynamics
simulations\cite{09.gR1,10.gR2,11.gR3,12.gR4}. The preferred direction
of the heat flux is observed from the wide to the narrow region.

Recently a new procedure is considered by Chang et al. in carbon and
boron nitride nanotubes\cite{13.Zhang}. They introduce mass gradient
along the axis of the nanotubes by covering external platinum compound
on the nanotubes. Since the platinum compound is almost thermal insulating
and the sp$^{2}$ bonds in nanotubes are much stronger than the bonds
between the fused external molecules, so Chang et al. idealize their
procedure as changing the mass of atoms in thermal conduction. Higher
thermal conductivity is observed when the heat flux runs from the
heavy to the light atoms and the rectification ratio is about 2-7\%.
Similar thermal rectification effect is also observed in 1D mass-graded
Fermi-Pasta-Ulam (FPU) $\beta$ chain and 1D single carbon chain\cite{08.rect05.LIbaowen,14.1d carbon}.
Influence of anharmonicity is surmised to explain the rectification
effect. Later Alaghemandi et al. have studied thermal conductivity
of mass-graded carbon nanotubes by RNEMD (reverse nonequilibrium molecular
dynamics) simulations\cite{15.mp1,16.mp2,17.mp3}. In their simulations,
the atomic mass of carbon atoms is gradually increased from 12 to
300 along the axis of the tube. The carbon atoms interact via harmonic
radial, angular and torsion potentials with constant force constants.
Much higher thermal conductivity is observed when the heat flux runs
from the light to the heavy atoms and the rectification ratio is about
20-50\%. The rectification effect is explained as the strong coupling
between the longitudinal and transverse modes in the carbon nanotubes\cite{17.mp3}. 

Since the structure and thermal property of graphene are similar to
carbon nanotube, one might expect the similar thermal rectification
in mass-graded graphene. However in this letter, we demonstrate that
mass-graded graphene flakes reveal no thermal rectification effect
by NEMD (nonequilibrium molecular dynamics) simulations. Here we refer
to graphene flakes with infinite width in thermal conduction process.
Since the geometric deformation along the width is avoided by using
the periodic boundary condition, it is used to approach the simulation
condition with infinite width. Different heat flux and mass gradient
are investigated to confirm the generality of the result. Furthermore,
we also study the finite wide graphene flakes by using free boundary
condition. A weak rectification effect is observed and the rectification
ratio is evidently approaching zero by increasing the width. Based
upon the numerical results, we discuss the mechanism leading to the
absence of thermal rectification.%
\begin{figure}
\centering{}\includegraphics[scale=0.14]{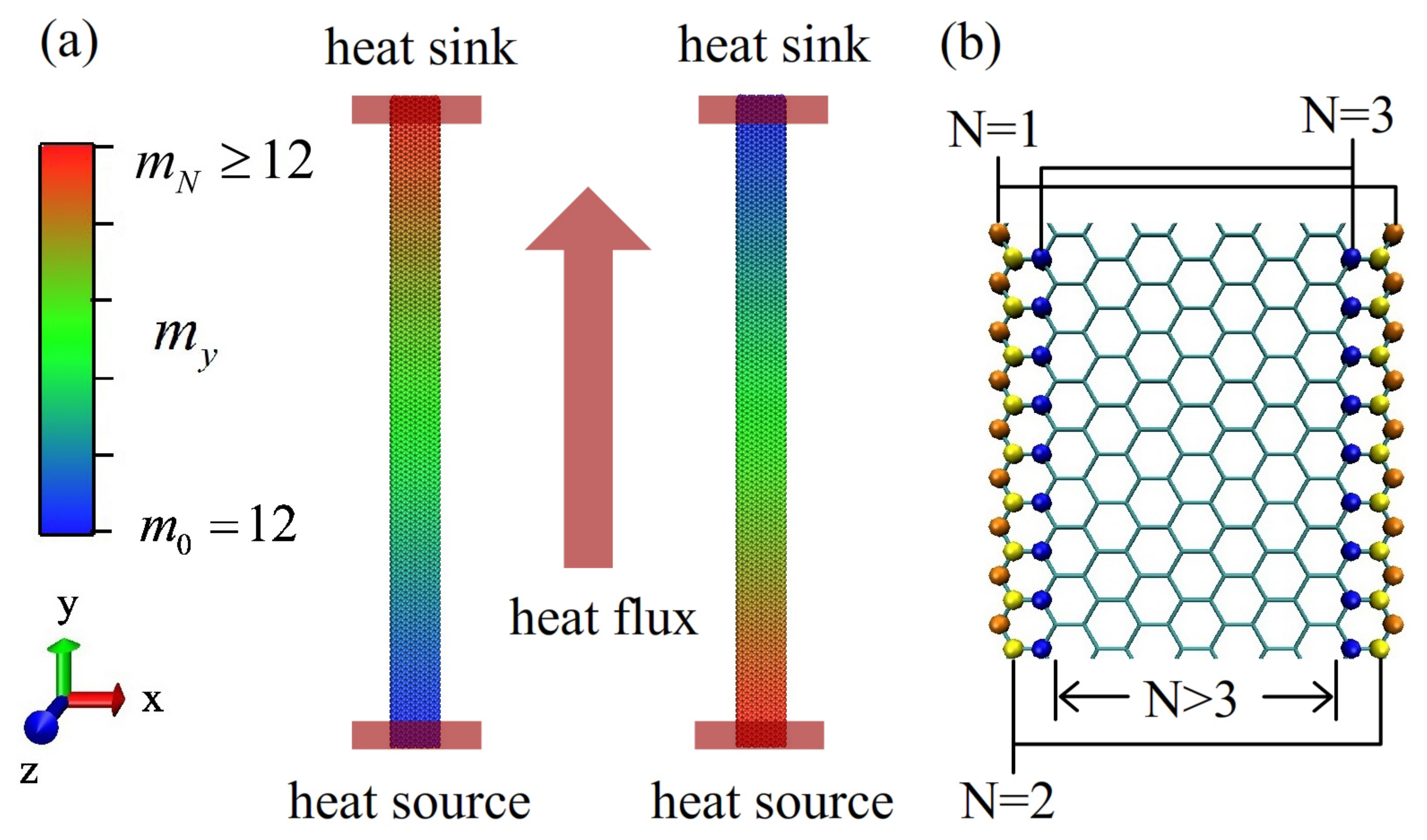}\caption{(Color online) (a) Schematic of the mass-graded graphene flakes. Heat
flux runs from the heat source to the heat sink. The atomic mass of
the carbon atoms varies from $m_{0}$ to $m_{N}$ according to Eq.
(1). Thus the first one is labeled as the $m_{0}-m{}_{N}$ graphene
flake and the second one is labeled as the $m_{N}-m{}_{0}$ graphene
flake. (b) A part of the graphene flake is amplified to show the edges
along the x-axis. $N=1$ are the outmost edge atoms which are drawn
in orange. $N=2$ edge atoms are drawn in yellow. $N=3$ edge atoms
are drawn in blue. $N>3$ are the inside atoms and only their bonds
are drawn. }

\end{figure}

We carry out the simulations in two graphene flakes with zig-zag edges
along the x-axis and armchair edges along the y-axis. The graphene
flakes are shown in Fig. 1(a). They are both 288 \foreignlanguage{american}{$\textrm{\AA}$}
long and 20 \foreignlanguage{american}{$\textrm{\AA}$} wide. The
heat sources and heat sinks are covered by red boxes. The outmost
edges of the heat sources/sinks are frozen. It corresponds to fixed
boundary conditions in the y-axis. Opposite mass gradients are implemented
in the two graphene flakes. The atomic mass of the carbon atoms between
the heat sources and the heat sinks along the y-axis is set as: 

\begin{equation}
m_{y}=\frac{m_{N}-m_{0}}{y_{N}-y_{0}}(y-y_{0})+m_{0}\end{equation}

\noindent Here $m{}_{0}=12$, $m_{N}\geqslant12$. We label the first
graphene flake in Fig. 1(a) as the $m_{0}-m{}_{N}$ graphene flake.
Its atomic mass varies from $m_{0}$ to $m_{N}$. Similarly we label
the second one as the $m_{N}-m{}_{0}$ graphene flake. The heat flux
runs along the $m_{N}-m{}_{0}$ graphene flake is equivalent to the
reversed heat flux runs along the $m_{0}-m{}_{N}$ graphene flake.
Thermal conduction process is investigated by imposing the same heat
flux along the two graphene flakes. It is much more convenient later
to compare the temperature profiles since the heat sources and heat
sinks are in the same direction. When using periodic boundary condition
along the x-axis, the edge atoms and the inside atoms are the same.
However when using free boundary condition along the x-axis, the edge
atoms would be less bound than the inside atoms. They have different
geometric surroundings. It is shown in Fig. 1(b), the edge atoms are
labelled as $N=1$, 2, 3 and the inside atoms are labeled as $N>3$.
We use the adaptive intermolecular reactive empirical bond-order (AIREBO)
potential\cite{18.rebo} as implemented in the LAMMPS\cite{19.LAMMPS}
code in the simulations. The REBO term simulates the anharmonic valence-bonded
C-C interactions in graphene. The bond energy consists of a repulsive
and attractive part:

\begin{equation}
E(r_{ij})=V_{R}(r_{ij})-b_{ij}V_{A}(r_{ij})\end{equation}

\noindent Here $V_{R}(r_{ij})=(1+\frac{Q}{r_{ij}})Ae^{-\alpha r_{ij}}$,$V_{A}(r_{ij})={\displaystyle \sum_{n=1,3}B_{n}e^{-\beta_{n}r_{ij}}}$,
$r_{ij}$ is the distance between the carbon atoms, $b_{ij}$ is a
function of the local coordination and bond angles for the $i$th
and $j$th atoms, $A$, $Q$, $B_{n}$, $\beta$ are parameters which
have been fitted according to carbon systems and can be found in the
original paper\cite{18.rebo}. Equations of motions are integrated
with velocity Verlet algorithm with the minimum timestep $\bigtriangleup t=0.25$
fs. %
\begin{figure}
\centering{}\includegraphics[scale=0.2]{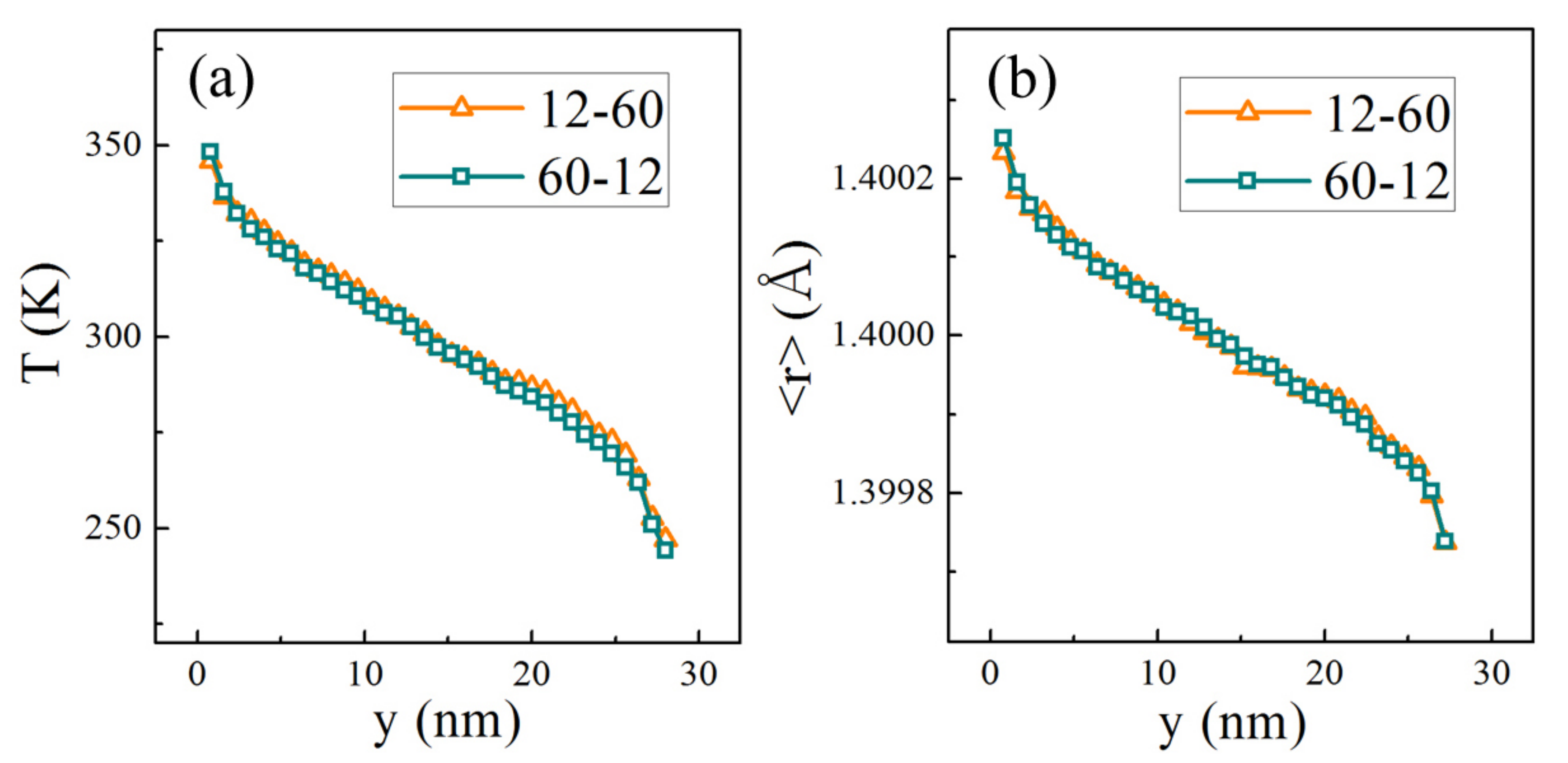}\caption{(Color online) Here $m_{N}/m_{0}=5$ and $J=0.35$ eV/ps are implemented.
(a) The typical temperature profiles. The $12-60$ and $60-12$ graphene
flakes exhibit the same temperature distribution. (b) The typical
bond length profiles. Since periodic boundary condition is used in
the x-axis, there is no difference between the edge atoms and the
inside atoms. So the bond length only varies along the y-axis. The
$12-60$ and $60-12$ graphene flakes also exhibit the same bond length
distribution.}

\end{figure}

First we study the thermal conduction of the two graphene flakes by
using periodic boundary condition along the x-axis. The graphene flakes
are equilibrated at a constant temperature $T=300$ K in the Nose-Hoover
thermostat by 0.75 ns. After that the heat flux is imposed. It is
realized by the energy and momentum conserving velocity rescaling
algorithm developed by Jude and Jullien\cite{20.FLUX,21.nanofluids,22.Interface2}.
By rescaling atomic velocities at each time step $dt$, specific amount
of kinetic energy $dE$ is added in the heat source and removed in
the heat sink respectively. The heat flux can be calculated by $J=dE/dt$.
We divide the graphene flakes by several 8 \foreignlanguage{american}{$\textrm{\AA}$}
long slabs to obtain the temperature profiles. The local temperature
in each slab is calculated from the averaged kinetic energy of the
carbon atoms. We average the temperature profiles over 100 ps after
the heat flux is imposed. The nonequilibrium simulation process covers
3 ns. Thermal conductivity $\kappa$ is obtained by the Fourier\textquoteright{}s
law:

\begin{equation}
\kappa=-\frac{J/A}{\Delta T/\Delta L}\end{equation}
Here $J$ is the heat flux, $A$ is the cross section of the heat
transfer defined by the width and thickness of the graphene flakes
(1.4 \foreignlanguage{american}{$\textrm{\AA}$} is considered as
the thickness), $\Delta T/\Delta L$ is the temperature gradient.
The reported data represents the steady state over the last 1000 ps.
The standard error of statistical uncertainties is within 5\%. Similarly,
in order to understand the geometric deformation in thermal conduction,
the bond length ($<r>$) profiles are also obtained. The bond length
describes the average distance of a carbon atom between its three
neighbors.

In Fig. 2(a) we show the typical temperature profiles of the two graphene
flakes. $m_{N}/m_{0}=5$ and $J=0.35$ eV/ps are implemented. The
two temperature profiles are the same which indicates the heat flux
runs equivalently without preferred direction. Their thermal conductivities
$\kappa=75$ W/mK are the same. Thus unlike mass-graded carbon nanotubes,
mass-graded graphene flakes reveal no obvious thermal rectification
effect even a large mass gradient is implemented. 

In Fig. 2(b) we show the associated bond length profiles. There is
no obvious difference between the two profiles. It indicates the geometric
deformation is also insensitive to the direction of the mass gradient.
Furthermore, a positive thermal expansion is observed along the longitudinal
direction. The bond lengths near the heat sources are larger than
those near the heat sinks. The average bond length of all the carbon
atoms is still equal to 1.4 \foreignlanguage{american}{$\textrm{\AA}$}
because the widths of the graphene flakes are unchanged by using periodic
boundary condition in the x-axis.%
\begin{figure}
\centering{}\includegraphics[scale=0.2]{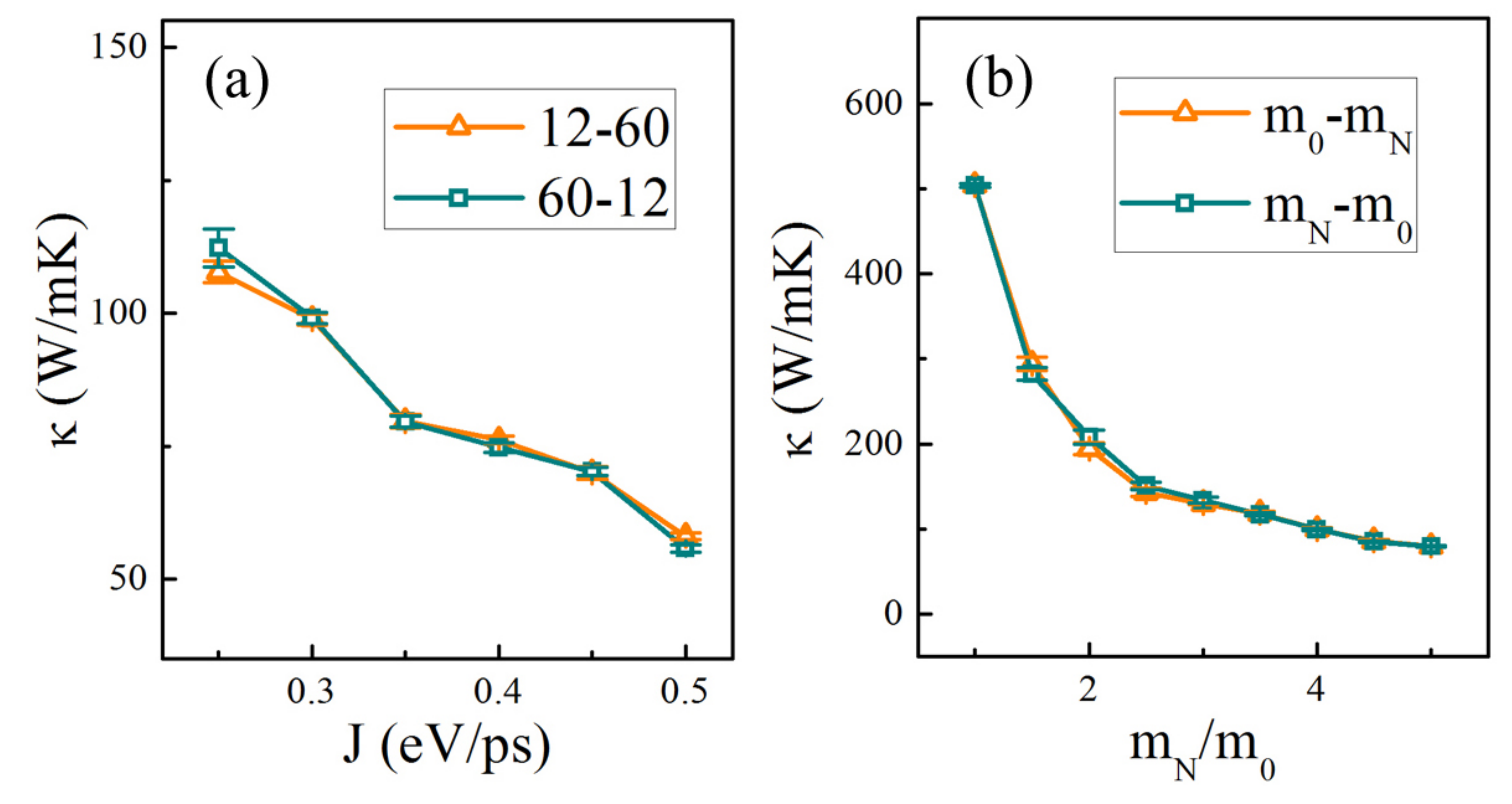}\caption{(Color online) (a) Thermal conductivity $\kappa$ vs heat flux $J$.
Here the mass gradient $m_{N}/m_{0}=5$ is unchanged. For $J\geqslant0.3$
eV/ps, the standard error of $\kappa$ is less than 2\% and the difference
of $\kappa$ is also less than 2\%. For $J=0.25$ eV/ps, the standard
error for the $12-16$ graphene flake is 2.2\% and the standard error
for the $60-12$ graphene flake is 3.9\%. The difference of $\kappa$
is 4.6\%. (b) Thermal conductivity $\kappa$ vs mass gradient $m_{N}/m_{0}$.
Here the heat flux $J=0.35$ eV/ps is unchanged. For $m_{N}/m_{0}=1$,
it stands for the graphene flake without mass gradient. The difference
of $\kappa$ is also within 5\%. }

\end{figure}

In order to confirm there is no thermal rectification effect, different
heat flux and mass gradient are applied. First we keep the mass gradient
$m_{N}/m_{0}=5$ unchanged and vary the heat flux $J$ from 0.25 to
0.5 eV/ps. In Fig. 3(a) we show the dependence of thermal conductivity
upon the heat flux. If the heat flux $J\geqslant0.3$ eV/ps, the difference
of thermal conductivities between the two graphene flakes is less
than 2\%. The effect of thermal fluctuation is stronger for small
heat flux value, so it would lower the computation certainty. Thus
for $J=0.25$ eV/ps, the difference of thermal conductivities becomes
4.6\% which is still very small. The result suggests there is no thermal
rectification effect by varying the heat flux. It is different from
mass-graded carbon nanotubes. Their thermal conductivity difference
would be as large as 20-50\%\cite{16.mp2,17.mp3}. 

Second we keep the heat flux $J=0.35$ eV/ps unchanged and vary the
mass gradient $m_{N}/m_{0}$ from 1 to 5. In Fig. 3(b) we show the
dependence of thermal conductivity upon the mass gradient. The thermal
conductivity difference is still within 5\%. The result suggests there
is no thermal rectification effect by varying the mass gradient. It
is also different from mass-graded carbon nanotubes. Their thermal
conductivity difference increases with the mass gradient\cite{16.mp2,17.mp3}.
Furthermore, the thermal conductivity of the graphene flakes decrease
dramatically with the mass gradient. If there is no mass gradient,
the thermal conductivity is 505 W/mK. When the mass gradient $m_{N}/m_{0}=5$
is implemented, it is reduced to 75 W/mK. It is only about 15\% of
the original value. So it provides a possible route to tune the thermal
behavior of graphene by modulating the mass gradient. Two methods
could be considered to implement the mass gradient. One is to load
external heavy and thermal insulating molecules. It has been widely
accepted as an idealized method to implement mass gradient in carbon
nanotubes\cite{13.Zhang,15.mp1,16.mp2}. The other one is to use different
ratio of isotope substitutions. Simulations of carbon nanotubes\cite{23.cnt iso}
and graphene\cite{24.grahene iso0 ,25.graphene iso1} of isotope defects
have already predicted the thermal conductivity reduction. The method
is demonstrated possible in experiment by chemical vapor deposition
growth of graphene on metal\cite{26.iso exp}.%
\begin{figure}
\centering{}\includegraphics[scale=0.2]{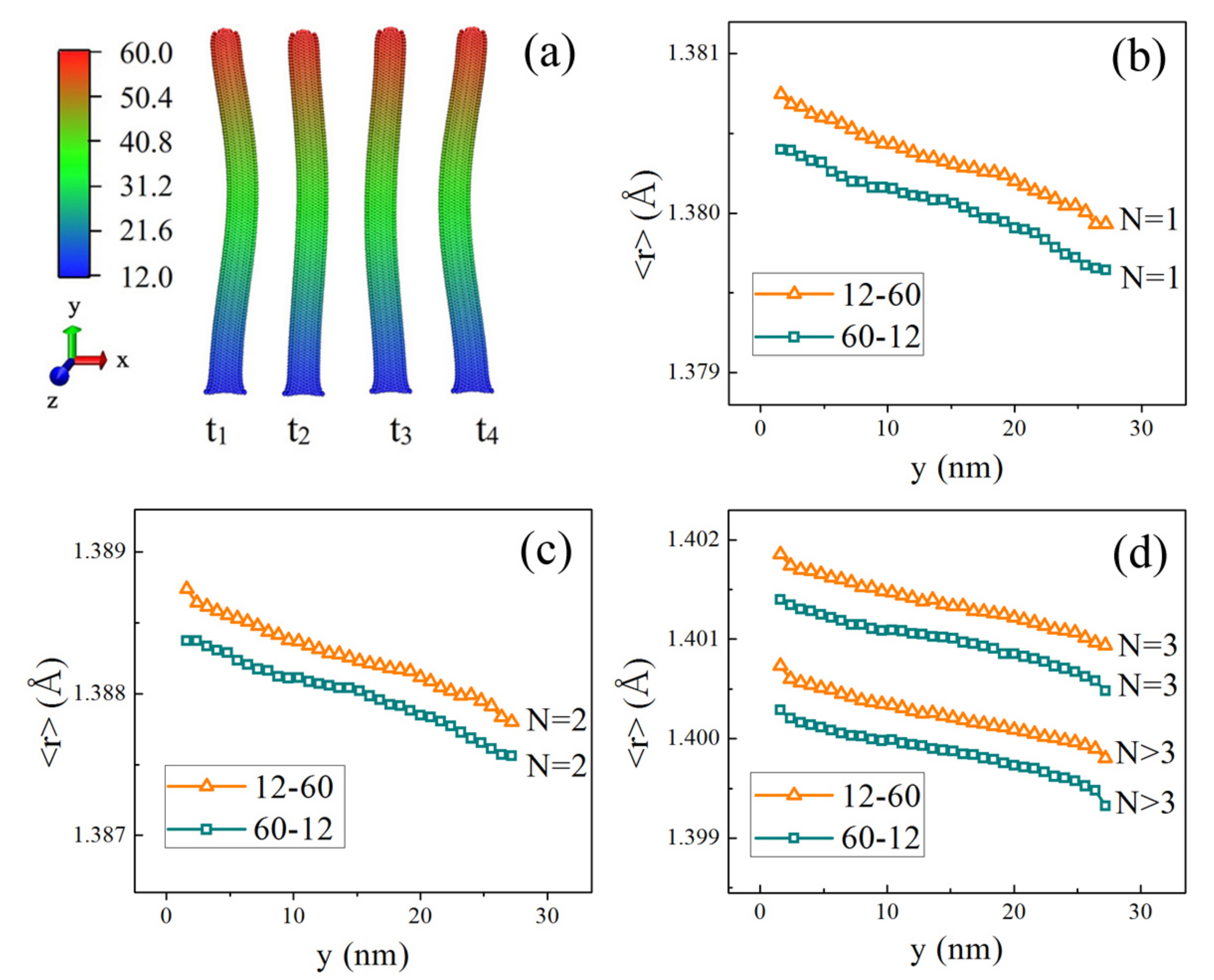}\caption{(Color online) (a) Thermal curvature in the $12-60$ graphene flake.
Time interval between two neighboring event is 25 ps ($t_{2}-t_{1}=t_{3}-t_{2}=t_{4}-t_{3}=25$
ps). The associated displacements are amplified 80 times for clarity.
{[}(b), (c) and (d){]} The average bond length profiles of the edge
atoms and inside atoms. N=1, 2, 3 are the edge atoms and N>3 are the
inside atoms in both graphene flakes. }

\end{figure}

To explain why rectification effect is not observed, we need to consider
the influence of asymmetric shape, anharmonicity and mode coupling.
They are the underlying mechanisms involved in thermal rectification.
It is known that different influence of those attributes in the physical
properties of 1D single chains and quasi-1D nanotubes, raises quite
different rectification effect\cite{16.mp2,17.mp3}. Thus it is also
interesting to understand how those attributes contribute in mass-graded
graphene which is a 2D system.

First we consider the influence of asymmetric shape. The geometric
deformation along the width is avoided by using the periodic boundary
condition in the x-axis. It is used to approach the simulation condition
that the infinite width is considered. As shown in Fig. 2(b), the
geometric deformations are the same in both graphene flakes. So the
geometric deformations bring no rectification effect. Later we shall
discuss the case when the finite width is considered.

Second we consider the influence of anharmonicity. Low-frequency modes
contributing predominantly to thermal conduction can be generated
when the anharmonic part of the FPU potential is excited. Thus heat
flux runs preferentially from the heavy to the light atoms in 1D mass-graded
Fermi-Pasta-Ulam (FPU) $\beta$ chain\cite{08.rect05.LIbaowen}. To
explain why such effect is not observed in mass-graded graphene, we
postulate that in carbon systems anharmonicity is insufficient to
bring an obvious rectification. For example, in 1D carbon chain, only
by implementing an extremely large mass gradient ($m_{N}/m_{0}\geqslant$32),
such rectification effect might be confirmed\cite{14.1d carbon}.
Thus for the mass gradients used in our simulations ($m_{N}/m_{0}\leqslant$5),
the influence of anharmonicity would be too weak to be detected. Furthermore,
thermal conductivity is dramatically decreased with the mass gradient.
So the application of the extremely large mass gradient might be very
limited.

Third we consider the influence of mode coupling. The coupling between
the longitudinal and transverse modes is more efficient when the light
atoms are placed at high temperature regions. Thus heat flux runs
preferentially from the light to the heavy atoms in mass-graded carbon
nanotubes. However it is also known that the mode coupling within
a 2D plane is much less strong than mode coupling in a bend topology,
such as quasi-1D nanotubes. Unlike mass-graded carbon nanotubes, the
associated rectification effect is very small in mass-graded graphene
flakes which belong to the 2D system\cite{17.mp3}. So it is difficult
to distinguish the effect of mode coupling from the thermal fluctuation
in the MD simulations.%
\begin{figure}
\centering{}\includegraphics[scale=0.17]{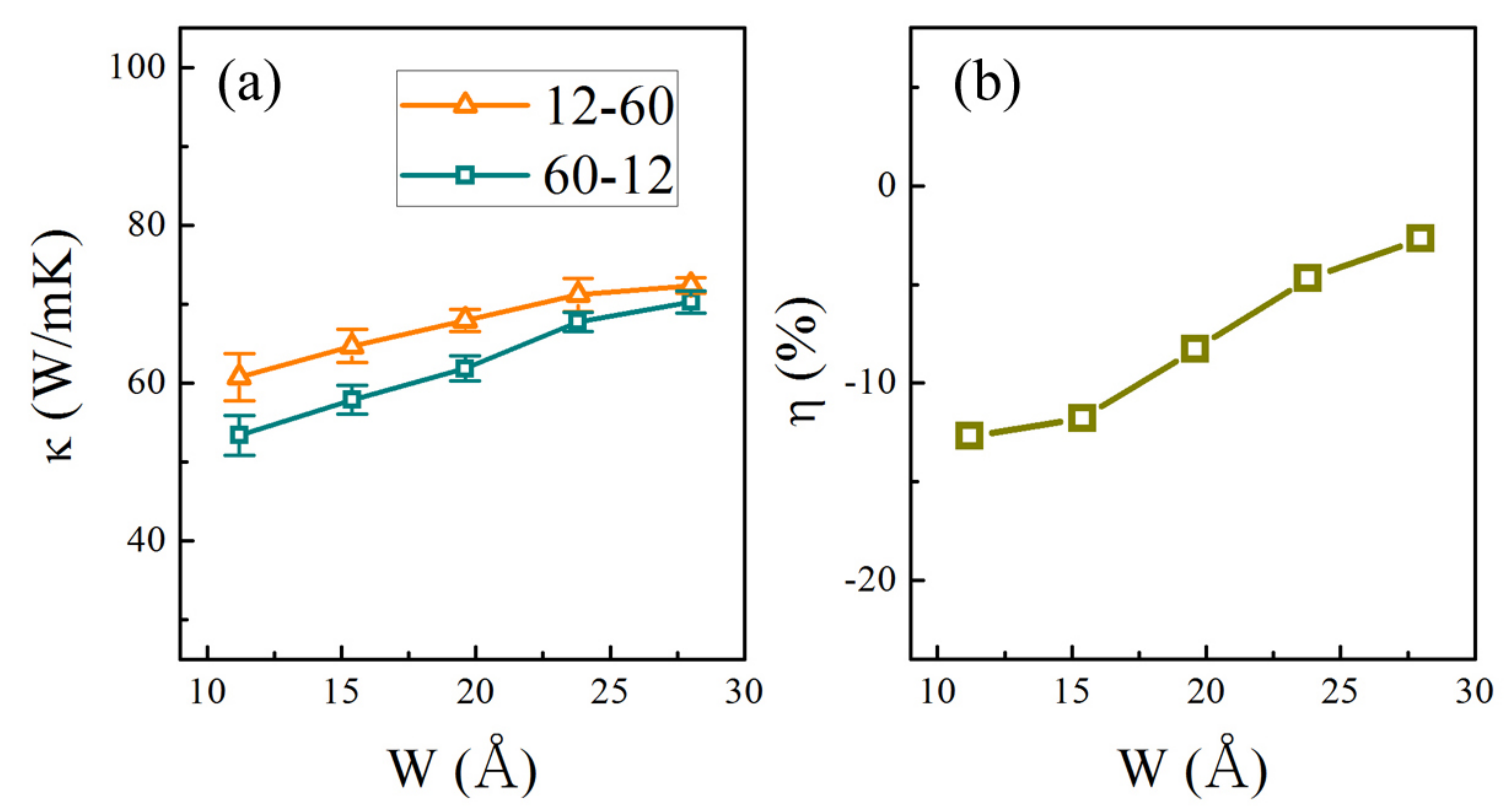}\caption{(Color online) (a) Thermal conductivity $\kappa$ vs width $W$. Here
$m_{N}/m_{0}=5$ and $J=0.35$ eV/ps are implemented. We only choose
the graphene flakes with the zig-zag edges along the x-axis. The graphene
flakes have the N=1, 2, 3 edge atoms and the N>3 inside atoms. The
fluctuation brought by thermal curvature and thermal expansion is
stronger if the width is small, so the computation uncertainty is
also greater. (b) Thermal rectification ratio $\eta$ vs the width
$W$. The average value of $\kappa$ in (a) is used to show the trend
thus no error bar is depicted.}

\end{figure}

By using periodic boundary condition in the x-axis, it corresponds
to the simulation condition that the infinite width is imitated. In
order to understand thermal conduction in finite wide mass-graded
graphene flakes, free boundary condition in the x-axis is considered.
Geometric deformation in the x-axis is possible in this case. $m_{N}/m_{0}=5$
and $J=0.35$ eV/ps are implemented. 

In Fig. 4(a) we show the thermal curvature in the $12-60$ graphene
flake. It is incessantly twisted in the thermal conduction process.
Similar thermal curvature is also observed in the $60-12$ graphene
flake. Thermal conductivity of the $12-60$ graphene flake is 67.8
W/mK. Thermal conductivity of the $60-12$ graphene flake is 61.4
W/mK. They are smaller than 75 W/mK in Fig. 2 where periodic boundary
condition is used. Similar reduction of thermal conductivity caused
by geometric deformation has been observed in graphene and carbon
nanotube recently\cite{27. weining,28.co.mp}.

In Fig. 4(b)-(d) we show the bond length profiles in the $12-60$
and $60-12$ graphene flakes. As shown in Fig. 1(b), when using free
boundary condition, the edge atoms are less bound with different geometric
surroundings. So their bond lengths are different from the inside
atoms. The bond lengths of the $N=1,2$ edges atoms are much smaller
than 1.4 \foreignlanguage{american}{$\textrm{\AA}$} which corresponds
to negative thermal expansion along the width. Meanwhile the bond
lengths of the $N=3$ edge atoms and the inside atoms are barely affected.
It indicates the negative thermal expansion occurs mostly along the
edge. The transverse negative thermal expansion in graphene has already
been observed in experiment and simulation\cite{29.Nat.Negative,30.PRB NEG}.

The bond length profiles in Fig. 4(b)-(d) also indicate asymmetric
geometric deformation occurs. The bond lengths $<r>$ in the $60-12$
graphene flake is smaller than $<r>$ in the $12-60$ graphene flake.
It is responsible for the observed weak thermal rectification effect.
The thermal rectification ratio $\eta$ is defined as\cite{13.Zhang,15.mp1,16.mp2,17.mp3}: 

\begin{equation}
\eta=\frac{\kappa_{m_{N}-m_{0}}-\kappa_{m_{0}-m_{N}}}{\kappa_{m_{0}-m_{N}}}\times100\%\end{equation}

\noindent Thus for the $12-60$ and $60-12$ graphene flakes, the
rectification ratio is $\eta=-9\%$. The negative sign of $\eta$
indicates heat flux runs preferentially from the light to the heavy
atoms. It is similar to mass-graded carbon nanotubes\cite{15.mp1,16.mp2,17.mp3}. 

In order to understand the influence of finite width, different width
$W$ is considered. Here we only choose the graphene flakes with two
zig-zag edges along the x-axis. Thus the chiralities of the graphene
flakes is unchanged. Free boundary condition in the x-axis is applied.
$m_{N}/m_{0}=5$ and $J=0.35$ eV/ps are implemented. In Fig. 5(a)
we show the dependence of thermal conductivity upon the width. The
result indicates higher thermal conductivity is obtained by placing
the heat source near the light atoms. In Fig. 5(b) we show the dependence
of the rectification ratio upon the width. We use the average value
of $\kappa$ in Fig. 5(a) to obtain $\eta$. The rectification ratio
is very small and evidently approaching 0 by increasing the width.
The result suggests the thermal rectification effect might be very
weak and cannot be observed in real application because the width
of graphene flake used in experiment is quantitatively large.

In summary, mass gradients in graphene flakes can not lead to thermal
rectification effect if the finite width effect is avoided by using
periodic boundary condition. Different heat flux and mass gradient
are considered to confirm the result. Meanwhile a weak rectification
effect is observed by using free boundary condition. However the rectification
ratio is evidently approaching zero by increasing the width. The result
implies that geometric deformation, anharmonicity and mode coupling
in mass-graded graphene are insufficient to bring rectification effect.
Additionally, mass gradient is demonstrated as an effective way in
tuning the thermal behavior of graphene. We hope our work shed light
on understanding the thermal conduction in mass-graded carbon systems
and designing graphene based thermal devices.

\acknowledgments We thank Jiao Wang, Yong Zhang and Dahai He for
helpful discussion and preparing of the manuscript. This work was
supported by National Natural Science Foundation of China(\#10775115
and \#10925525).

\end{document}